\documentclass[prl,preprint,superscriptaddress]{revtex4-1}
\usepackage{setspace}
\usepackage{amsfonts,amsmath,amssymb}
\usepackage{txfonts}
\usepackage{graphicx}
\usepackage{bm}
\usepackage{epstopdf}
\usepackage{verbatim}
\usepackage{units}
\usepackage[T1]{fontenc}
\usepackage[latin9]{inputenc}
\usepackage{amsbsy}
\usepackage{esint}
 
\begin{document}
 
\title{Nonmodal growth of the magnetorotational instability}

\author{J.~Squire}
\affiliation{Department of Astrophysical Sciences and Princeton Plasma Physics Laboratory, Princeton University, Princeton, NJ 08543}
\author{A.~Bhattacharjee}
\affiliation{Department of Astrophysical Sciences and Princeton Plasma Physics Laboratory, Princeton University, Princeton, NJ 08543}
\affiliation{Max Planck/Princeton Center for Plasma Physics, Princeton University, Princeton, NJ 08543}

\begin{abstract}
We analyze the linear growth of the magnetorotational instability (MRI) in the short time limit using nonmodal methods. Our findings are quite different to standard results, illustrating that shearing wave energy can grow at the maximum MRI rate, $-d\Omega/d \ln r,$ for \emph{any} choice of azimuthal and vertical wavelengths. In addition, by comparing the growth of shearing waves with static structures, we show that over short time-scales  shearing waves will always be dynamically more important than static structures in the ideal limit. By demonstrating that  fast linear growth is possible at all wavelengths, these results suggest that nonmodal linear physics could play a fundamental role in MRI turbulence.
\end{abstract}

\pacs{52.30.Cv,47.20.Ft,97.10.Gz}

\date{\today}

\maketitle

Since the seminal work of Balbus and Hawley \cite{Balbus:1991fi}, the magnetorotational instability (MRI) has emerged as promising explanation for the observed momentum transport in accretion disks. In particular, the nonlinear development of the instability has been shown to lead to sustained turbulence and dynamo action in both local shearing box (\emph{e.g.,} \cite{Hawley:1995gd,Kapyla:2011fj,Simon:2012dq}) and global (\emph{e.g.,} \cite{Sorathia:2012jq}) nonlinear simulations. Despite substantial concern about transport convergence with dissipation parameters \cite{Fromang:2007cg,Longaretti:2010ha} it seems that such results are relatively robust \cite{Simon:2012dq}, persisting both with and without an imposed magnetic field and somewhat independently of boundary conditions and background density profiles \cite{Kapyla:2011fj}. Nonetheless, even in the simplest local case, a thorough understanding of the nature of the turbulence and the dynamo mechanism is lacking (see, for instance \cite{Blackman:2012wd}). Several promising closure models and dynamo ideas (\emph{e.g.,} \cite{Pessah:2006kc,Lesur:2008fn,Vishniac:1997jo,Heinemann:2011gt,Riols:2013dk,Rincon:2008jd}) require further testing, and there has been less work on the nature of the turbulent cascade \cite{Lesur:2011jh} (if it even exists in the usual sense \cite{Bratanov:2013it,Fromang:2007cg}). The relevance of linear eigenmodes in these processes seems to have mostly been discounted (\emph{e.g.,} \cite{Longaretti:2010ha}), although there have been hints that linear waves advected by the mean flow (shearing waves, sometimes called spatial Fourier harmonics or Kelvin modes) have substantial dynamical importance \cite{Lesur:2008fn,Heinemann:2011gt}.

The study of the linear eigenmodes of a system is, in the most basic sense, an attempt to answer the following question: How much can the system grow in time and what initial conditions will maximize this growth? If said system is self-adjoint in a physically relevant norm, the eigenspectrum is certainly the best way to approach this problem; initializing with the most unstable eigenmode will maximize growth of the disturbance at all times. However, the question becomes more complex if the linear operator is not self-adjoint and \emph{nonmodal} effects become important \cite{Trefethen:2005wt}. In particular, the answer can depend enormously on the time at which one wishes to maximize the growth, and the eigenvalue result is correct only in the limit $t\rightarrow \infty$. If one studies growth over shorter times, not only can growth rates be very much larger than predicted with eigenvalue analysis, but the most relevant structures can look very different to the eigenmodes. Investigations in this vein have been particularly fruitful in fluid dynamics, where they have cleanly answered longstanding questions about subcritical transition to turbulence in spectrally stable systems \cite{SCHMID:2007bz}.

In this letter we approach the linear stability of the MRI from the nonmodal standpoint, studying the \emph{short time} growth of perturbations. The picture that emerges suggests that eigenmode and asymptotic shearing wave growth estimates can be quite misleading, since over shorter time-scales relevant to a turbulent situation the growth can be very different than in the long-time limit. In particular, we prove that shearing wave structures  (we include the axisymmetric mode as a special case of this) always grow faster over short time-scales than static (eigenmode-like) structures so long as dissipation is not too large. Interestingly,  the local ideal short time energy growth rate has the \emph{same} maximum value, $-d \Omega/d \ln r,$ regardless of the vertical and azimuthal wavelengths. We also show how such calculations can be extended to more general situations with a weakly spatially dependent shearing wave expansion, considering an incompressible global model motivated by liquid metal experiments \cite{Rudiger:2013vm}. Finally, we confirm these ideas numerically, demonstrating that the fastest growing non-axisymmetric linear structures in global domains are shearing waves, with growth rates many times larger than the most unstable eigenmodes. Such calculations establish a natural connection between global modes and local shearing waves, illustrating that in almost all situations a local calculation will give a more accurate indication of important moderate-time linear behavior than the global eigenmodes. 

The significance of transient growth for non-axisymmetric MRI has been recognized in many previous papers (\emph{e.g.,} 
\cite{Balbus:1992du,Papaloizou:1997vn,Johnson:2007wo,Mamatsashvili:2013gy,Salhi:2012hd,Brandenburg:2006ec}), which have mainly focused on the transience brought about by the time-dependent spatial structure of shearing waves. Our considerations focus on the non-normality of the linear operator and differ in several respects from prior works:
\begin{enumerate}
\item Curiously, it is generally assumed that shearing waves are the most relevant structures in local inquiries, while most global studies consider eigenmodes (but see \cite{Papaloizou:1997vn,Shtemler:2012ty}). We advocate that the dynamical relevance of each type of structure should be determined based on growth rates, since strongly amplified structures will dominate when starting from random initial conditions. With this in mind, we \emph{prove} (within the WKB approximation) that in almost all regimes, shearing waves grow faster on short time-scales. 
\item We find that the fastest short-time shearing wave growth occurs in a regime where previous analytic results \cite{Balbus:1992du,Johnson:2007wo,Mamatsashvili:2013gy} are not valid.
\item We find that transient growth can be significant even for axisymmetric modes in the local case (channel modes). 
\item We argue that both shearing waves and eigenmodes can be important in many situations. While shearing waves invariably grow faster over moderate time-scales, they can transition into an eigenmode as the radial wavenumber becomes large and continue growing (if the eigenmode is unstable). In this way the eventual decay of shearing waves at finite diffusivity is not physically important, even without consideration of nonlinearities. This viewpoint provides a natural connection between local and global non-axisymmetric modes.
\end{enumerate}  

\emph{Local Calculation.} Our starting point  is the local incompressible MHD equations in a rotating frame
\begin{align}
\frac{\partial\bm{u}}{\partial t}&+\left(\bm{u}\cdot\nabla\right)\bm{u}+2\Omega \bm{\hat{z}}\times \bm{u}=-\nabla p+\nabla\times\bm{B}\times\bm{B}\nonumber \\ 
&\;\;\;+2q \Omega^2 x \bm{\hat{x}}- \nabla \Phi+\bar{\nu}\nabla^{2}\bm{u},\nonumber \\
\frac{\partial\bm{B}}{\partial t}&+\left(\bm{u}\cdot\nabla\right)\bm{B}=\left(\bm{B}\cdot\nabla\right)\bm{u} + \bar{\eta}\nabla^2\bm{b},\nonumber\\
&\nabla\cdot\bm{u}=0,\;\;\;\nabla\cdot\bm{B}=0.\label{shearMHD}
\end{align}
These are obtained from the standard MHD equations with radial stratification by considering a small Cartesian volume (at $r_0$) co-rotating with the fluid at angular velocity $\Omega\left(r\right)\sim \Omega_0 r^{-q}$ \cite{Umurhan:2004fm}. In Eqs.~\eqref{shearMHD}, the directions $x,\,y,\,z$ correspond respectively to the radial, azimuthal and vertical directions in the disk. We have used dimensionless variables normalized by the length scale $L_0$ and the time-scale $1/\Omega$ -- as such $\Omega\equiv\Omega\left(r_0\right)=1$ in Eqs.~\eqref{shearMHD}. Since all parameters in our problem are of order $1$, the fluid and magnetic diffusivities, $\bar{\nu}$ and $\bar{\eta}$, are the inverses of the fluid and magnetic Reynolds numbers respectively.  The background velocity is azimuthal with linear shear in the radial direction, $\bm{u}_0=-q\Omega x\hat{\bm{y}}$, and the background magnetic field is taken to be constant, $\bm{B}_0=\left(0,B_{0y},B_{0z}\right)$. We linearize about this background and Fourier analyze in $y$ and $z$, denoting the respective wave-numbers $k_y$ and $k_z$. Changing to convenient Orr-Sommerfeld like variables \cite{Johnson:2007wo},
$u=u_x,\;B=B_x,\:\zeta=i k_z u_y-i k_y u_z,\;\eta=i k_z B_y-i k_y B_z$,
we are left with 4 coupled partial differential equations in $x$ and $t$.

The general idea of nonmodal growth calculations is to compute, for some chosen time, the initial conditions that lead to the maximum growth of the solution under the chosen norm. We use the energy of the perturbation,
$E=\int d\bm{x}\left( \left|\bm{u}\right|^2+\left|\bm{B}\right|^2\right),$ as the norm throughout this work, since it seems the most physically relevant choice \cite{SCHMID:2007bz}. For the sake of clarity, consider the general linear system 
$\partial U/\partial t=\mathcal{L}(t) U(t),$ with solution $U(t)=K(t)U(0)$. The maximum growth at $t$,
$G(t)=\max_{U(0)} \left\Vert K(t) U(0) \right\Vert_E^2/\left\Vert U(0) \right\Vert_E^2$
(where $\left\Vert U \right\Vert_E^2=U^{\dagger}\cdot M_E(t)\cdot U$ denotes the energy norm of $U$), can be calculated by changing to the 2-norm using the Cholesky decomposition 
\begin{equation}
\left\Vert U \right\Vert_E^2=U^{\dagger}\cdot M_E(t)\cdot U=U^{\dagger}\cdot F^{\dagger}(t)F(t)\cdot U= \left\Vert F(t) U \right\Vert_2^2,\label{F defn}
\end{equation}
and computing the largest singular value of the matrix $F(t)K(t)F^{-1}(0)$. 
For the analytic results presented in this letter, we compute the growth rate at $t=0^+$, 
$G^+_{max}=\max_{U(0)}\left. \left\Vert U(t) \right\Vert_E^{-2} \frac{d}{dt}\left\Vert U(t) \right\Vert_E^2 \right|_{t=0^+}.$ 
Note that for a self-adjoint system $G^+_{max}$ is simply (twice) the most unstable eigenvalue. Differentiating $K(t)$, changing to the 2-norm and defining $\Lambda=\left. F \mathcal{L} F^{-1}+\partial_t F F^{-1} \right|_{t=0}$, we obtain the result
\begin{equation}
G^+_{max}=\lambda_{max}\left(\Lambda+\Lambda^{\dagger}\right)\label{G+result},
\end{equation}
where $\lambda_{max}$ denotes the largest eigenvalue  \cite{Farrell:1996bc}.

Motivated by the ubiquitous occurrence of shearing waves in simulations, we wish to compare the growth of shearing structures with eigenmodes. Noting that the defining characteristic of an eigenmode is that its wavenumber is constant in time, we consider these at a given $x$ value using a WKB approximation and term these static waves. While caution is advised in attempting to predict stability using such methods \cite{Knobloch:1992vk}, here we are simply comparing static and shearing growth at a \emph{given} $k_x$. Thus, subtle issues regarding the choice of $k_x$ relevant to an eigenmode are alleviated and we make no claim that these approximations are a substitute for the solution of the $x$ dependent problem (but see \cite{Blokland:2005id}). Note that in \emph{both} cases (static and shearing) the growth calculation is nonmodal; we  insert an ansatz for the spatial form of the disturbance to better understand the structures that will appear in an $x$ dependent nonmodal solution. The static equations can easily be derived  to lowest order by inserting the WKB ansatz $f\!\left(x,t\right)\sim f\!\left(t\right)\, e^{i k_x x}$, and substituting $\frac{\partial}{\partial t}\rightarrow \frac{\partial}{\partial t}- i \bm{u}_0 k_y=\frac{\partial}{\partial t}+ i q x k_y$ (this simply shifts the spectrum without changing growth rates). 
The shearing wave equations are locally exact 
\footnote{Note that the local shearing wave equations are actually nonlinearly valid due to fortuitous cancellations in Eqs.~\eqref{shearMHD} upon insertion of the shearing wave ansatz.} 
and are derived by inserting $f\!\left(x,t\right)=f\!\left(t\right)\exp\left(i q k_y  \left(t-t_0\right) x\right)$ for each independent variable. We obtain
\begin{equation}
\frac{\partial}{\partial t}U\!\left(t\right)=
\left(
\begin{array}{cccc}
 -\bar{\nu}  k^2-2 \Xi q k_x k_y/k^2 & -2 i k_z/k^2 & i F & 0 \\
 i (q-2)k_z & -k^2 \bar{\nu}  & 0 & i F \\
 i F & 0 & -k^2 \bar{\eta}  & 0 \\
 0 & i F & -i q k_z & -k^2 \bar{\eta}  \\
\end{array}
\right) \cdot U\!\left(t\right).\label{SWeqns}
\end{equation}
where $\Xi =0\text{ or }1$ for static and shearing waves respectively, $F=k_y B_{0y}+k_z B_{0z}$, $k^2=k_x^2+k_y^2+k_z^2$, $U\!\left(t\right) =\left(u,\zeta,B,\eta\right)$ and we have used $\Omega=1$. For the shearing waves, the equations are time-dependent since $k_x=q k_y \left(t-t_0\right)$. Solving for the eigenvalues of Eqs.~\eqref{SWeqns} with $k_x=k_y=0$ leads to the standard MRI dispersion relation \cite{Balbus:1991fi}.

Converting the energy norm into $\left(u,\zeta,B,\eta\right)$ variables gives the inner product \\$F=\sqrt{2\pi^2 \left(k_y^2+k_z^2\right)^{-1}}\mathrm{diag}\!\left(k,1,k,1\right)$, where $\mathrm{diag}\left(\,\right)$ denotes the diagonal matrix [see Eq.~\eqref{F defn}]. Using Eq.~\eqref{G+result} we obtain the remarkably simple results:
\begin{equation}
G^+_{max}=
\max\begin{cases}
q\frac{k_z}{k}-2\bar{\nu} k^2 & \\
q\frac{k_z}{k}-2\bar{\eta} k^2 & 
\end{cases}\label{StatG+}
\end{equation}
for the static waves (with $\max\!\left\{\colon \!\right.$ denoting the maximum of the two functions), and
\begin{equation}
G^+_{max}=\max\begin{cases}
q\left(\frac{1}{k}\sqrt{k_z^2+\frac{k_x^2 k_y^2}{k^2}}-\frac{k_x k_y}{k^2}\right)-2\bar{\nu} k^2 & \\
q\left(\frac{1}{k}\sqrt{k_z^2+\frac{k_x^2 k_y^2}{k^2}}+\frac{k_x k_y}{k^2}\right)-2\bar{\eta} k^2 &
\end{cases}\label{SWG+}
\end{equation}
for the shearing wave solutions, with $k$, $k_x$ evaluated at $t=0$. 

Consider first the ideal limit of Eqs.~\eqref{StatG+} and \eqref{SWG+}, $\bar{\nu}=\bar{\eta}=0$. We see that at all wave-numbers the shearing wave can grow faster than a static structure (or as fast at $k_x=0$ where they are identical).  In addition, the shearing wave growth rate has maxima at $k_x(0)=\pm k_y$, at which the growth is $q \Omega$, \emph{i.e.,} the maximum eigenvalue of the MRI, reached when $k_y=k_x=0,\:k_z=1/B_{0z} \sqrt{15/16}$. Thus, in the ideal limit, the MRI can have the same growth rate, $q \Omega$, for any choice of $k_y,\:k_z$, so long as  the shearing wave initial condition satisfies $k_x(0)=\pm k_y$. We note that \emph{all} channel mode perturbations ($k_x=k_y=0$) grow at the same rate $q \Omega$, showing that even this most basic of MRI modes can grow transiently when $k_z\ne1/B_{0z} \sqrt{15/16}$. This transient growth is a real physical effect; indeed, in the ideal limit any axisymmetric perturbation involving $B_x$ can grow arbitrarily large through simple advection.  At all wave numbers, the initial conditions to obtain $G^+_{max}$ are either purely hydrodynamic or purely magnetic.  Of course, these pure modes will quickly become mixed under time-evolution due to coupling terms in Eqs.~\eqref{SWeqns}.
Unsurprisingly, adding dissipation alters this result. In particular, static waves can grow \emph{faster} than shearing waves at sufficiently high wave-numbers when $\text{Pm}=\bar{\nu}/\bar{\eta}\ne 1$, with $\mathrm{Pm}>1$ ($\mathrm{Pm}<1$) causing static structures to dominate for $k_x(0)< 0$ ($k_x(0)>0$).

\emph{Inclusion of global effects.} We can extend this result to situations in which aspects of the local approximation may not hold (see \emph{e.g.,} \cite{Pessah:2005cp,Kirillov:2013db}) by starting our analysis from the standard MHD equations in cylindrical co-ordinates \cite{Bondeson:1987bq} and considering shearing and static waves with weak dependence on the radial co-ordinate. Motivated by liquid metal experiments \cite{Rudiger:2013vm,Kirillov:2013db}, here we consider the incompressible MHD equations at constant density with the velocity profile $\bm{u}_0=U_{0\theta} r^{-q +1}\hat{\bm{\theta}}$ and the magnetic field profile $\bm{B}_0=B_{0\theta} r^{2 \text{Rb} +1}\hat{\bm{\theta}}+B_{0z}\hat{\bm{z}}$. The extension of the technique to more complex stratifications and compressibility \cite{Pessah:2005cp,Salhi:2012hd} is straightforward. Non-dimensionalizing the equations and  Fourier analyzing in $\theta$ and $z$, we obtain a system of 8 linear PDEs in $r$ and $t$. These are reduced to four equations with the variable transformation, $u=u_r,\;B=B_r,\;\zeta=i k_z u_{\theta}-i \frac{m}{r} u_z,\;\eta=i k_z B_{\theta}-i \frac{m}{r} B_z$, where $m$ and $k_z$ are the azimuthal and vertical wave numbers.

The static wave equations are obtained in much the same way as for the local case, by inserting the ansatz $f\!\left(r,t\right)\sim f\!\left(t\right)\, e^{i k_r r}$ and assuming $\left(k_r r,\:k_z r,\:m\right)\sim1/\epsilon$, $\left(\bar{\nu},\:\bar{\eta}\right)\sim \epsilon^2$ to obtain a set of ODEs in time \footnote{These ordering assumptions arise from assuming the solution varies faster than the background equilibrium. If either $\bar{\nu}$ or $\bar{\eta}$ are very large one may wish to alter these, leading to slightly different shearing wave equations}. Similarly, the shearing wave equations are obtained by assuming a shearing wave envelope that varies slowly in the $r$ direction. To lowest order, they can be straightforwardly derived by inserting the ansatz $f\!\left(r,t\right)\sim f\!\left(t\right)\,\exp \left(-i \frac{m}{r} U_0 r^{-q+1} (t-t_0)\right)$ and making the same ordering assumptions as for the static case. 
After non-dimensionalizing variables using the length-scale $r$ and the time-scale $1/\Omega(r)$, one obtains
\begin{align}
&\partial_t U= \nonumber\\
&\left(
\begin{array}{cccc}
 -k^2\bar{\nu}-2 q \Xi \frac{m k_r}{k^2} & -2 i k_z/k^2 & i F\left(r\right) & 2 i k_z B_{az}/k^2 \\
 i (q-2) k_z & -k^2 \bar{\nu}  & 2 i (\text{Rb}+1) k_z B_{az} & i F\left(r\right) \\
 i F\left(r\right) & 0 & -k^2 \bar{\eta}  & 0 \\
 -2 i \text{Rb} k_z B_{az } & i F\left(r\right) & -i q k_z & -k^2 \bar{\eta}  \\
\end{array}
\right)\cdot U.\label{GlobalSWeqns}
\end{align}
Here $U=\left(u,\zeta,B,\eta\right)$, $\Xi=1$ or 0 for shearing waves and static waves respectively, $B_{az}=B_{0\theta}r^{2\text{Rb}+1}$, $F(r)=k_z B_{0 z}+m B_{az}$, wave numbers ($k_r$, $k_z$) have been scaled by $r$ and ($\bar{\nu}$, $\bar{\eta}$) have been scaled by $r^2 \Omega(r)$. In the static case we have substituted $\frac{\partial}{\partial t}\rightarrow \frac{\partial}{\partial t}- i \bm{u}_0 m/r$ (as for the local calculation)  and for the shearing wave, $k_r=q U_0 m r^{-q }(t-t_0)$. While all variables in Eqs.~\eqref{GlobalSWeqns} technically depend on both $r$ and $t$, the dependence on $r$ is parametric. The static version ($\Xi=0$) of Eqs.~\eqref{GlobalSWeqns} is very similar to the dispersion relation given in \cite{Kirillov:2013db}, aside from slight differences in how the azimuthal wavenumber $m$ appears in the dissipation terms. Note that Eqs.~\eqref{GlobalSWeqns} reduce to Eqs.~\eqref{SWeqns} in the "local" limit \cite{Umurhan:2004fm}.

Applying the same procedure as earlier to calculate the $t=0^+$ growth rates leads to
\begin{align}
G^+_{max}=&\pm\left[\left(k^2\left(\bar{\eta}-\bar{\nu}\right)- \Xi q \frac{k_r m}{k^2}\right)^2+4 B_{az} \left(1+\text{Rb}\right)^2\frac{k_z^2}{k^2}\right]^\frac{1}{2}\nonumber\\
+&\frac{q}{k} \sqrt{k_z^2+\Xi \frac{m^2k_r^2}{k^2}}-k^2\left(\bar{\eta}+\bar{\nu}\right),\label{Global G+}
\end{align}
with the $\pm$ chosen to obtain the maximum value of $G^+_{max}$. Note that $\left| r\, \partial_r f\right|^2\approx r^2 \left| \partial_r f\right|^2$ has been applied in the energy norm used to calculate Eq.~\eqref{Global G+}, in keeping with approximations used earlier.  Eq.~\eqref{Global G+} demonstrates that the fundamental results presented earlier are essentially unchanged by the addition of field curvature effects, as well as illustrating the importance of shearing waves in flows with more complex shear profiles.  The extra terms in the global equations change the maximum of $G^+_{max}$ with respect to $k_x(0)$, and the MRI can grow faster than $q\Omega$ for strong $B_{0\theta}$. It is interesting that for the very large $\bar{\eta}$ characteristic of liquid metal experiments there is a large regime (for $k_x(0)>0$) where static structures grow faster than shearing waves.

\begin{figure}
\begin{centering}

\includegraphics[width=1.0\columnwidth]{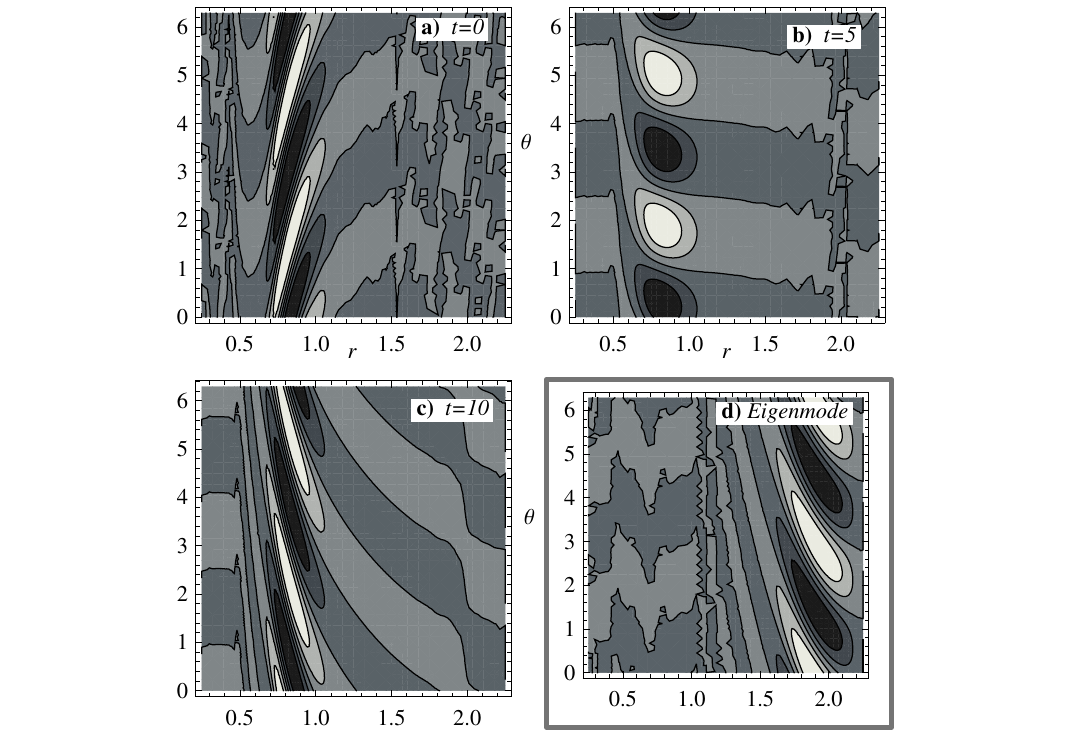}
\caption{(a)-(c) Time evolution of the spatial structure of the radial magnetic field perturbation in $\left(r,\theta \right)$ that maximizes energy amplification at $t_M=10$.  The global parameters are: $m=2,\:k_z=15,\:U_0=1,\:B_{0z}=1/30,\:B_{az}=0,\:\bar{\nu}=\bar{\eta}=1/10000$. White and black regions show positive and negative values respectively. (d, boxed) Spatial structure of the most unstable eigenmode for the same parameters. \label{Figure1}}

\end{centering}
\end{figure}
\begin{figure}
\begin{centering}

\includegraphics[width=1.0\columnwidth]{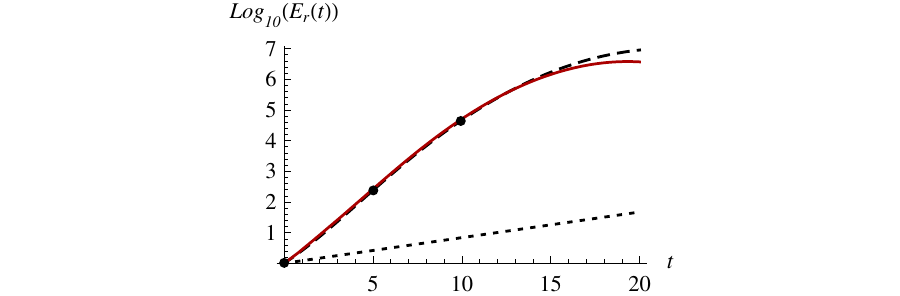}
\caption{Energy amplification, $E_r\left(t\right)\equiv E\left(r=1,t\right)/E\left(r=1,0\right)$, of the global nonmodal solution (dashed), the most unstable eigenmode (dotted), and the solution of the local shearing wave equations [Eqs.~\eqref{SWeqns}] (solid, red). Parameters are the same as Fig.~\ref{Figure1}. The black dots illustrate the time frames shown in Fig.~\ref{Figure1}.\label{Figure2}}

\end{centering}
\end{figure}

\emph{Nonmodal growth in a global domain.} We now illustrate the relevance of nonmodal growth in global domains with boundaries. As well as demonstrating that shearing waves [Eqs.~\eqref{SWeqns} and \eqref{GlobalSWeqns}] often have greater applicability than eigenmodes to the global linear problem, the nonmodal standpoint provides a concrete connection between global modes and shearing box dynamics.  We solve the incompressible MHD equations in cylindrical co-ordinates with hard-wall boundaries, linearized about the background flow velocity $\bm{u}_0=U_{0\theta} r^{-1/2}\hat{\bm{\theta}}$. We discretize radially with Chebyshev polynomials on the domain $r=0.25\rightarrow2.25$, and consider a single azimuthal and vertical wavenumber. While it would be more realistic to include density stratification and compressibility in such a model, the general conclusions are unaltered by the addition of such effects. As outlined in Ref.~\cite{SCHMID:2007bz}, the nonmodal calculation solves for the initial conditions that maximize the energy amplification by some chosen time $t_M$.

The time evolution of the spatial structure of this nonmodal solution for a weakly non-axisymmetric mode (with $t_M=10$) is illustrated in Fig.~\ref{Figure1}(a)-(c), with the most unstable eigenmode shown for comparison [Fig.~\ref{Figure1}(d)]. Note the strong resemblance of the nonmodal structure to a shearing wave as well as its localization far from the boundary (around $r\approx 1$) in stark contrast to the eigenmode. To demonstrate the dynamical relevance of the shearing wave equations as opposed to the global eigenmode, in Fig.~\ref{Figure2} we compare the energy growth of the nonmodal structure (at $r=1$) with that of the solution to Eqs.~\eqref{GlobalSWeqns} \footnote{The lack of an azimuthal field renders Eqs.~\eqref{SWeqns} and \eqref{GlobalSWeqns} identical in this example} and the most unstable eigenmode. The parameters in Eqs.~\eqref{GlobalSWeqns} are taken from the global parameters (with $t_0=4.6$ to match the global nonmodal solution), and the initial conditions are calculated using the nonmodal technique, maximizing energy growth at $t=10$. The most obvious feature in Fig.~\ref{Figure2} is that the nonmodal solution grows many times faster than the eigenmode, showing its greater dynamical relevance. In addition, we see that the shearing wave approximation is very accurate, with the only difference coming at late times when the global solution departs from a pure shearing wave due to the finite resistivity. Evidently, the strong flow shear necessitates the use of nonmodal techniques in MRI stability calculations, while consideration of shearing waves allows the extension of local stability methods to these situations with surprising accuracy.

\emph{Conclusions.} We have analyzed the short time growth rate of the MRI from a nonmodal standpoint. By comparing the growth of shearing  and static waves, we prove that shearing structures always dominate in the ideal limit and that the peak growth rates are identical to those of the axisymmetric channel mode at all scales. Of course, this result contains no information about the length of time over which the short time growth can persist, and thus the overall amplification of a given mode over finite times. Indeed, since the growth rates are completely independent of magnetic field, more information is certainly needed to consider a quasi-linear mechanism for the accretion disk dynamo \footnote{In fact, a similar result holds for short time growth in the equivalent sheared rotating hydrodynamic system, where it seems that there is never a transition to the turbulent state}.  Most important is probably the provision for growth over finite time-scales (\emph{e.g.,} Fig.~\ref{Figure2}) or with driving noise \cite{Farrell:1994wh}, as well as the effects of spatial inhomogeneity in the background fields \cite{Lesur:2008fn}. Aside from such problems, the results are suggestive about the character of the turbulence in shearing MHD systems. In particular, the strong linear amplification over short times at all dynamically relevant scales supports the idea that MRI turbulence should not exhibit a well defined inertial range \cite{Fromang:2007cg,Bratanov:2013it}. Given this, might many of the turbulence properties be well understood by considering primarily linear physics? In any case no matter how applicable such concepts might turn out to be, it seems clear that an over-reliance on eigenvalue and dispersion relation analyses can lead to incorrect growth predictions in many regimes. 

We extend thanks to Dr. Jeremy Goodman for enlightening discussion. This work was supported by Max Planck/Princeton Center for Plasma Physics and  U.S. DOE (DE-AC02-09CH11466).


\begin{thebibliography}{38}%
\makeatletter
\providecommand \@ifxundefined [1]{%
 \@ifx{#1\undefined}
}%
\providecommand \@ifnum [1]{%
 \ifnum #1\expandafter \@firstoftwo
 \else \expandafter \@secondoftwo
 \fi
}%
\providecommand \@ifx [1]{%
 \ifx #1\expandafter \@firstoftwo
 \else \expandafter \@secondoftwo
 \fi
}%
\providecommand \natexlab [1]{#1}%
\providecommand \enquote  [1]{``#1''}%
\providecommand \bibnamefont  [1]{#1}%
\providecommand \bibfnamefont [1]{#1}%
\providecommand \citenamefont [1]{#1}%
\providecommand \href@noop [0]{\@secondoftwo}%
\providecommand \href [0]{\begingroup \@sanitize@url \@href}%
\providecommand \@href[1]{\@@startlink{#1}\@@href}%
\providecommand \@@href[1]{\endgroup#1\@@endlink}%
\providecommand \@sanitize@url [0]{\catcode `\\12\catcode `\$12\catcode
  `\&12\catcode `\#12\catcode `\^12\catcode `\_12\catcode `\%12\relax}%
\providecommand \@@startlink[1]{}%
\providecommand \@@endlink[0]{}%
\providecommand \url  [0]{\begingroup\@sanitize@url \@url }%
\providecommand \@url [1]{\endgroup\@href {#1}{\urlprefix }}%
\providecommand \urlprefix  [0]{URL }%
\providecommand \Eprint [0]{\href }%
\providecommand \doibase [0]{http://dx.doi.org/}%
\providecommand \selectlanguage [0]{\@gobble}%
\providecommand \bibinfo  [0]{\@secondoftwo}%
\providecommand \bibfield  [0]{\@secondoftwo}%
\providecommand \translation [1]{[#1]}%
\providecommand \BibitemOpen [0]{}%
\providecommand \bibitemStop [0]{}%
\providecommand \bibitemNoStop [0]{.\EOS\space}%
\providecommand \EOS [0]{\spacefactor3000\relax}%
\providecommand \BibitemShut  [1]{\csname bibitem#1\endcsname}%
\let\auto@bib@innerbib\@empty
\bibitem [{\citenamefont {Balbus}\ and\ \citenamefont
  {Hawley}(1991)}]{Balbus:1991fi}%
  \BibitemOpen
  \bibfield  {author} {\bibinfo {author} {\bibfnamefont {S.~A.}\ \bibnamefont
  {Balbus}}\ and\ \bibinfo {author} {\bibfnamefont {J.~F.}\ \bibnamefont
  {Hawley}},\ }\href@noop {} {\bibfield  {journal} {\bibinfo  {journal}
  {Astrophys. J.}\ }\textbf {\bibinfo {volume} {376}},\ \bibinfo {pages} {214}
  (\bibinfo {year} {1991})}\BibitemShut {NoStop}%
\bibitem [{\citenamefont {Hawley}\ \emph {et~al.}(1995)\citenamefont {Hawley},
  \citenamefont {Gammie},\ and\ \citenamefont {Balbus}}]{Hawley:1995gd}%
  \BibitemOpen
  \bibfield  {author} {\bibinfo {author} {\bibfnamefont {J.~F.}\ \bibnamefont
  {Hawley}}, \bibinfo {author} {\bibfnamefont {C.~F.}\ \bibnamefont {Gammie}},
  \ and\ \bibinfo {author} {\bibfnamefont {S.~A.}\ \bibnamefont {Balbus}},\
  }\href@noop {} {\bibfield  {journal} {\bibinfo  {journal} {Astrophys. J.}\
  }\textbf {\bibinfo {volume} {440}},\ \bibinfo {pages} {742} (\bibinfo {year}
  {1995})}\BibitemShut {NoStop}%
\bibitem [{\citenamefont {K{\"a}pyl{\"a}}\ and\ \citenamefont
  {Korpi}(2011)}]{Kapyla:2011fj}%
  \BibitemOpen
  \bibfield  {author} {\bibinfo {author} {\bibfnamefont {P.~J.}\ \bibnamefont
  {K{\"a}pyl{\"a}}}\ and\ \bibinfo {author} {\bibfnamefont {M.~J.}\
  \bibnamefont {Korpi}},\ }\href@noop {} {\bibfield  {journal} {\bibinfo
  {journal} {Mon. Not. R. Astron. Soc.}\ }\textbf {\bibinfo {volume} {413}},\
  \bibinfo {pages} {901} (\bibinfo {year} {2011})}\BibitemShut {NoStop}%
\bibitem [{\citenamefont {Simon}\ \emph {et~al.}(2012)\citenamefont {Simon},
  \citenamefont {Beckwith},\ and\ \citenamefont {Armitage}}]{Simon:2012dq}%
  \BibitemOpen
  \bibfield  {author} {\bibinfo {author} {\bibfnamefont {J.~B.}\ \bibnamefont
  {Simon}}, \bibinfo {author} {\bibfnamefont {K.}~\bibnamefont {Beckwith}}, \
  and\ \bibinfo {author} {\bibfnamefont {P.~J.}\ \bibnamefont {Armitage}},\
  }\href@noop {} {\bibfield  {journal} {\bibinfo  {journal} {Mon. Not. R.
  Astron. Soc.}\ }\textbf {\bibinfo {volume} {422}},\ \bibinfo {pages} {2685}
  (\bibinfo {year} {2012})}\BibitemShut {NoStop}%
\bibitem [{\citenamefont {Sorathia}\ \emph {et~al.}(2012)\citenamefont
  {Sorathia}, \citenamefont {Reynolds}, \citenamefont {Stone},\ and\
  \citenamefont {Beckwith}}]{Sorathia:2012jq}%
  \BibitemOpen
  \bibfield  {author} {\bibinfo {author} {\bibfnamefont {K.~A.}\ \bibnamefont
  {Sorathia}}, \bibinfo {author} {\bibfnamefont {C.~S.}\ \bibnamefont
  {Reynolds}}, \bibinfo {author} {\bibfnamefont {J.~M.}\ \bibnamefont {Stone}},
  \ and\ \bibinfo {author} {\bibfnamefont {K.}~\bibnamefont {Beckwith}},\
  }\href@noop {} {\bibfield  {journal} {\bibinfo  {journal} {Astrophys. J.}\
  }\textbf {\bibinfo {volume} {749}},\ \bibinfo {pages} {189} (\bibinfo {year}
  {2012})}\BibitemShut {NoStop}%
\bibitem [{\citenamefont {Fromang}\ and\ \citenamefont
  {Papaloizou}(2007)}]{Fromang:2007cg}%
  \BibitemOpen
  \bibfield  {author} {\bibinfo {author} {\bibfnamefont {S.}~\bibnamefont
  {Fromang}}\ and\ \bibinfo {author} {\bibfnamefont {J.}~\bibnamefont
  {Papaloizou}},\ }\href@noop {} {\bibfield  {journal} {\bibinfo  {journal}
  {Astron. Astrophys.}\ }\textbf {\bibinfo {volume} {476}},\ \bibinfo {pages}
  {1113} (\bibinfo {year} {2007})}\BibitemShut {NoStop}%
\bibitem [{\citenamefont {Longaretti}\ and\ \citenamefont
  {Lesur}(2010)}]{Longaretti:2010ha}%
  \BibitemOpen
  \bibfield  {author} {\bibinfo {author} {\bibfnamefont {P.~Y.}\ \bibnamefont
  {Longaretti}}\ and\ \bibinfo {author} {\bibfnamefont {G.}~\bibnamefont
  {Lesur}},\ }\href@noop {} {\bibfield  {journal} {\bibinfo  {journal} {Astron.
  Astrophys.}\ }\textbf {\bibinfo {volume} {516}},\ \bibinfo {pages} {51}
  (\bibinfo {year} {2010})}\BibitemShut {NoStop}%
\bibitem [{\citenamefont {Blackman}(2012)}]{Blackman:2012wd}%
  \BibitemOpen
  \bibfield  {author} {\bibinfo {author} {\bibfnamefont {E.~G.}\ \bibnamefont
  {Blackman}},\ }\href@noop {} {\bibfield  {journal} {\bibinfo  {journal}
  {Physica Scripta}\ }\textbf {\bibinfo {volume} {86}},\ \bibinfo {pages}
  {058202} (\bibinfo {year} {2012})}\BibitemShut {NoStop}%
\bibitem [{\citenamefont {Pessah}\ \emph {et~al.}(2006)\citenamefont {Pessah},
  \citenamefont {Chan},\ and\ \citenamefont {Psaltis}}]{Pessah:2006kc}%
  \BibitemOpen
  \bibfield  {author} {\bibinfo {author} {\bibfnamefont {M.~E.}\ \bibnamefont
  {Pessah}}, \bibinfo {author} {\bibfnamefont {C.-k.}\ \bibnamefont {Chan}}, \
  and\ \bibinfo {author} {\bibfnamefont {D.}~\bibnamefont {Psaltis}},\
  }\href@noop {} {\bibfield  {journal} {\bibinfo  {journal} {Phys. Rev. Lett.}\
  }\textbf {\bibinfo {volume} {97}},\ \bibinfo {pages} {221103} (\bibinfo
  {year} {2006})}\BibitemShut {NoStop}%
\bibitem [{\citenamefont {Lesur}\ and\ \citenamefont
  {Ogilvie}(2008)}]{Lesur:2008fn}%
  \BibitemOpen
  \bibfield  {author} {\bibinfo {author} {\bibfnamefont {G.}~\bibnamefont
  {Lesur}}\ and\ \bibinfo {author} {\bibfnamefont {G.~I.}\ \bibnamefont
  {Ogilvie}},\ }\href@noop {} {\bibfield  {journal} {\bibinfo  {journal} {Mon.
  Not. R. Astron. Soc.}\ }\textbf {\bibinfo {volume} {391}},\ \bibinfo {pages}
  {1437} (\bibinfo {year} {2008})}\BibitemShut {NoStop}%
\bibitem [{\citenamefont {Vishniac}\ and\ \citenamefont
  {Brandenburg}(1997)}]{Vishniac:1997jo}%
  \BibitemOpen
  \bibfield  {author} {\bibinfo {author} {\bibfnamefont {E.~T.}\ \bibnamefont
  {Vishniac}}\ and\ \bibinfo {author} {\bibfnamefont {A.}~\bibnamefont
  {Brandenburg}},\ }\href@noop {} {\bibfield  {journal} {\bibinfo  {journal}
  {Astrophys. J.}\ }\textbf {\bibinfo {volume} {475}},\ \bibinfo {pages} {263}
  (\bibinfo {year} {1997})}\BibitemShut {NoStop}%
\bibitem [{\citenamefont {Heinemann}\ \emph {et~al.}(2011)\citenamefont
  {Heinemann}, \citenamefont {McWilliams},\ and\ \citenamefont
  {Schekochihin}}]{Heinemann:2011gt}%
  \BibitemOpen
  \bibfield  {author} {\bibinfo {author} {\bibfnamefont {T.}~\bibnamefont
  {Heinemann}}, \bibinfo {author} {\bibfnamefont {J.~C.}\ \bibnamefont
  {McWilliams}}, \ and\ \bibinfo {author} {\bibfnamefont {A.~A.}\ \bibnamefont
  {Schekochihin}},\ }\href@noop {} {\bibfield  {journal} {\bibinfo  {journal}
  {Phys. Rev. Lett.}\ }\textbf {\bibinfo {volume} {107}},\ \bibinfo {pages}
  {255004} (\bibinfo {year} {2011})}\BibitemShut {NoStop}%
\bibitem [{\citenamefont {Riols}\ \emph {et~al.}(2013)\citenamefont {Riols},
  \citenamefont {Rincon}, \citenamefont {Cossu}, \citenamefont {Lesur},
  \citenamefont {Longaretti}, \citenamefont {Ogilvie},\ and\ \citenamefont
  {Herault}}]{Riols:2013dk}%
  \BibitemOpen
  \bibfield  {author} {\bibinfo {author} {\bibfnamefont {A.}~\bibnamefont
  {Riols}}, \bibinfo {author} {\bibfnamefont {F.}~\bibnamefont {Rincon}},
  \bibinfo {author} {\bibfnamefont {C.}~\bibnamefont {Cossu}}, \bibinfo
  {author} {\bibfnamefont {G.}~\bibnamefont {Lesur}}, \bibinfo {author}
  {\bibfnamefont {P.~Y.}\ \bibnamefont {Longaretti}}, \bibinfo {author}
  {\bibfnamefont {G.~I.}\ \bibnamefont {Ogilvie}}, \ and\ \bibinfo {author}
  {\bibfnamefont {J.}~\bibnamefont {Herault}},\ }\href@noop {} {\bibfield
  {journal} {\bibinfo  {journal} {J. Fluid Mech.}\ }\textbf {\bibinfo {volume}
  {731}},\ \bibinfo {pages} {1} (\bibinfo {year} {2013})}\BibitemShut {NoStop}%
\bibitem [{\citenamefont {Rincon}\ \emph {et~al.}(2008)\citenamefont {Rincon},
  \citenamefont {Ogilvie}, \citenamefont {Proctor},\ and\ \citenamefont
  {Cossu}}]{Rincon:2008jd}%
  \BibitemOpen
  \bibfield  {author} {\bibinfo {author} {\bibfnamefont {F.}~\bibnamefont
  {Rincon}}, \bibinfo {author} {\bibfnamefont {G.~I.}\ \bibnamefont {Ogilvie}},
  \bibinfo {author} {\bibfnamefont {M.~R.~E.}\ \bibnamefont {Proctor}}, \ and\
  \bibinfo {author} {\bibfnamefont {C.}~\bibnamefont {Cossu}},\ }\href@noop {}
  {\bibfield  {journal} {\bibinfo  {journal} {Astron. Nach.}\ }\textbf
  {\bibinfo {volume} {329}},\ \bibinfo {pages} {750} (\bibinfo {year}
  {2008})}\BibitemShut {NoStop}%
\bibitem [{\citenamefont {Lesur}\ and\ \citenamefont
  {Longaretti}(2011)}]{Lesur:2011jh}%
  \BibitemOpen
  \bibfield  {author} {\bibinfo {author} {\bibfnamefont {G.}~\bibnamefont
  {Lesur}}\ and\ \bibinfo {author} {\bibfnamefont {P.~Y.}\ \bibnamefont
  {Longaretti}},\ }\href@noop {} {\bibfield  {journal} {\bibinfo  {journal}
  {Astron. Astrophys.}\ }\textbf {\bibinfo {volume} {528}},\ \bibinfo {pages}
  {17} (\bibinfo {year} {2011})}\BibitemShut {NoStop}%
\bibitem [{\citenamefont {Bratanov}\ \emph {et~al.}(2013)\citenamefont
  {Bratanov}, \citenamefont {Jenko}, \citenamefont {Hatch},\ and\ \citenamefont
  {Wilczek}}]{Bratanov:2013it}%
  \BibitemOpen
  \bibfield  {author} {\bibinfo {author} {\bibfnamefont {V.}~\bibnamefont
  {Bratanov}}, \bibinfo {author} {\bibfnamefont {F.}~\bibnamefont {Jenko}},
  \bibinfo {author} {\bibfnamefont {D.~R.}\ \bibnamefont {Hatch}}, \ and\
  \bibinfo {author} {\bibfnamefont {M.}~\bibnamefont {Wilczek}},\ }\href@noop
  {} {\bibfield  {journal} {\bibinfo  {journal} {Phys. Rev. Lett.}\ }\textbf
  {\bibinfo {volume} {111}},\ \bibinfo {pages} {075001} (\bibinfo {year}
  {2013})}\BibitemShut {NoStop}%
\bibitem [{\citenamefont {Trefethen}\ and\ \citenamefont
  {Embree}(2005)}]{Trefethen:2005wt}%
  \BibitemOpen
  \bibfield  {author} {\bibinfo {author} {\bibfnamefont {L.~N.}\ \bibnamefont
  {Trefethen}}\ and\ \bibinfo {author} {\bibfnamefont {M.}~\bibnamefont
  {Embree}},\ }\href@noop {} {\emph {\bibinfo {title} {{Spectra and
  Pseudospectra}}}},\ The Behavior of Nonnormal Matrices and Operators\
  (\bibinfo  {publisher} {Princeton University Press},\ \bibinfo {year}
  {2005})\BibitemShut {NoStop}%
\bibitem [{\citenamefont {Schmid}(2007)}]{SCHMID:2007bz}%
  \BibitemOpen
  \bibfield  {author} {\bibinfo {author} {\bibfnamefont {P.~J.}\ \bibnamefont
  {Schmid}},\ }\href@noop {} {\bibfield  {journal} {\bibinfo  {journal} {Annu.
  Rev. Fluid Mech.}\ }\textbf {\bibinfo {volume} {39}},\ \bibinfo {pages} {129}
  (\bibinfo {year} {2007})}\BibitemShut {NoStop}%
\bibitem [{\citenamefont {R{\"u}diger}\ \emph {et~al.}(2013)\citenamefont
  {R{\"u}diger}, \citenamefont {Hollerbach},\ and\ \citenamefont
  {Kitchatinov}}]{Rudiger:2013vm}%
  \BibitemOpen
  \bibfield  {author} {\bibinfo {author} {\bibfnamefont {G.}~\bibnamefont
  {R{\"u}diger}}, \bibinfo {author} {\bibfnamefont {R.}~\bibnamefont
  {Hollerbach}}, \ and\ \bibinfo {author} {\bibfnamefont {L.~L.}\ \bibnamefont
  {Kitchatinov}},\ }\href@noop {} {\emph {\bibinfo {title} {{Magnetic Processes
  in Astrophysics}}}},\ Theory, Simulations, Experiments\ (\bibinfo
  {publisher} {John Wiley {\&} Sons},\ \bibinfo {year} {2013})\BibitemShut
  {NoStop}%
\bibitem [{\citenamefont {Balbus}\ and\ \citenamefont
  {Hawley}(1992)}]{Balbus:1992du}%
  \BibitemOpen
  \bibfield  {author} {\bibinfo {author} {\bibfnamefont {S.~A.}\ \bibnamefont
  {Balbus}}\ and\ \bibinfo {author} {\bibfnamefont {J.~F.}\ \bibnamefont
  {Hawley}},\ }\href@noop {} {\bibfield  {journal} {\bibinfo  {journal}
  {Astrophys. J.}\ }\textbf {\bibinfo {volume} {400}},\ \bibinfo {pages} {610}
  (\bibinfo {year} {1992})}\BibitemShut {NoStop}%
\bibitem [{\citenamefont {Papaloizou}\ and\ \citenamefont
  {Terquem}(1997)}]{Papaloizou:1997vn}%
  \BibitemOpen
  \bibfield  {author} {\bibinfo {author} {\bibfnamefont {J.~C.~B.}\
  \bibnamefont {Papaloizou}}\ and\ \bibinfo {author} {\bibfnamefont
  {C.}~\bibnamefont {Terquem}},\ }\href@noop {} {\bibfield  {journal} {\bibinfo
   {journal} {Mon. Not. R. Astron. Soc.}\ }\textbf {\bibinfo {volume} {287}},\
  \bibinfo {pages} {771} (\bibinfo {year} {1997})}\BibitemShut {NoStop}%
\bibitem [{\citenamefont {Johnson}(2007)}]{Johnson:2007wo}%
  \BibitemOpen
  \bibfield  {author} {\bibinfo {author} {\bibfnamefont {B.~M.}\ \bibnamefont
  {Johnson}},\ }\href@noop {} {\bibfield  {journal} {\bibinfo  {journal}
  {Astrophys. J.}\ }\textbf {\bibinfo {volume} {660}},\ \bibinfo {pages} {1375}
  (\bibinfo {year} {2007})}\BibitemShut {NoStop}%
\bibitem [{\citenamefont {Mamatsashvili}\ \emph {et~al.}(2013)\citenamefont
  {Mamatsashvili}, \citenamefont {Chagelishvili}, \citenamefont {Bodo},\ and\
  \citenamefont {Rossi}}]{Mamatsashvili:2013gy}%
  \BibitemOpen
  \bibfield  {author} {\bibinfo {author} {\bibfnamefont {G.~R.}\ \bibnamefont
  {Mamatsashvili}}, \bibinfo {author} {\bibfnamefont {G.~D.}\ \bibnamefont
  {Chagelishvili}}, \bibinfo {author} {\bibfnamefont {G.}~\bibnamefont {Bodo}},
  \ and\ \bibinfo {author} {\bibfnamefont {P.}~\bibnamefont {Rossi}},\
  }\href@noop {} {\bibfield  {journal} {\bibinfo  {journal} {Mon. Not. R.
  Astron. Soc.}\ }\textbf {\bibinfo {volume} {435}},\ \bibinfo {pages} {2552}
  (\bibinfo {year} {2013})}\BibitemShut {NoStop}%
\bibitem [{\citenamefont {Salhi}\ \emph {et~al.}(2012)\citenamefont {Salhi},
  \citenamefont {Lehner}, \citenamefont {Godeferd},\ and\ \citenamefont
  {Cambon}}]{Salhi:2012hd}%
  \BibitemOpen
  \bibfield  {author} {\bibinfo {author} {\bibfnamefont {A.}~\bibnamefont
  {Salhi}}, \bibinfo {author} {\bibfnamefont {T.}~\bibnamefont {Lehner}},
  \bibinfo {author} {\bibfnamefont {F.}~\bibnamefont {Godeferd}}, \ and\
  \bibinfo {author} {\bibfnamefont {C.}~\bibnamefont {Cambon}},\ }\href@noop {}
  {\bibfield  {journal} {\bibinfo  {journal} {Physical Review E}\ }\textbf
  {\bibinfo {volume} {85}},\ \bibinfo {pages} {026301} (\bibinfo {year}
  {2012})}\BibitemShut {NoStop}%
\bibitem [{\citenamefont {Brandenburg}\ and\ \citenamefont
  {Dintrans}(2006)}]{Brandenburg:2006ec}%
  \BibitemOpen
  \bibfield  {author} {\bibinfo {author} {\bibfnamefont {A.}~\bibnamefont
  {Brandenburg}}\ and\ \bibinfo {author} {\bibfnamefont {B.}~\bibnamefont
  {Dintrans}},\ }\href@noop {} {\bibfield  {journal} {\bibinfo  {journal}
  {Astron. Astrophys.}\ }\textbf {\bibinfo {volume} {450}},\ \bibinfo {pages}
  {437} (\bibinfo {year} {2006})}\BibitemShut {NoStop}%
\bibitem [{\citenamefont {Shtemler}\ \emph {et~al.}(2012)\citenamefont
  {Shtemler}, \citenamefont {Mond},\ and\ \citenamefont
  {Liverts}}]{Shtemler:2012ty}%
  \BibitemOpen
  \bibfield  {author} {\bibinfo {author} {\bibfnamefont {Y.~M.}\ \bibnamefont
  {Shtemler}}, \bibinfo {author} {\bibfnamefont {M.}~\bibnamefont {Mond}}, \
  and\ \bibinfo {author} {\bibfnamefont {E.}~\bibnamefont {Liverts}},\
  }\href@noop {} {\bibfield  {journal} {\bibinfo  {journal} {Mon. Not. R.
  Astron. Soc.}\ }\textbf {\bibinfo {volume} {421}},\ \bibinfo {pages} {700}
  (\bibinfo {year} {2012})}\BibitemShut {NoStop}%
\bibitem [{\citenamefont {Umurhan}\ and\ \citenamefont
  {Regev}(2004)}]{Umurhan:2004fm}%
  \BibitemOpen
  \bibfield  {author} {\bibinfo {author} {\bibfnamefont {O.~M.}\ \bibnamefont
  {Umurhan}}\ and\ \bibinfo {author} {\bibfnamefont {O.}~\bibnamefont
  {Regev}},\ }\href@noop {} {\bibfield  {journal} {\bibinfo  {journal} {Astron.
  Astrophys.}\ }\textbf {\bibinfo {volume} {427}},\ \bibinfo {pages} {855}
  (\bibinfo {year} {2004})}\BibitemShut {NoStop}%
\bibitem [{\citenamefont {Farrell}\ and\ \citenamefont
  {Ioannou}(1996)}]{Farrell:1996bc}%
  \BibitemOpen
  \bibfield  {author} {\bibinfo {author} {\bibfnamefont {B.~F.}\ \bibnamefont
  {Farrell}}\ and\ \bibinfo {author} {\bibfnamefont {P.~J.}\ \bibnamefont
  {Ioannou}},\ }\href@noop {} {\bibfield  {journal} {\bibinfo  {journal}
  {Journal of the Atmospheric Sciences}\ }\textbf {\bibinfo {volume} {53}},\
  \bibinfo {pages} {2041} (\bibinfo {year} {1996})}\BibitemShut {NoStop}%
\bibitem [{\citenamefont {Knobloch}(1992)}]{Knobloch:1992vk}%
  \BibitemOpen
  \bibfield  {author} {\bibinfo {author} {\bibfnamefont {E.}~\bibnamefont
  {Knobloch}},\ }\href@noop {} {\bibfield  {journal} {\bibinfo  {journal} {Mon.
  Not. R. Astron. Soc.}\ }\textbf {\bibinfo {volume} {255}},\ \bibinfo {pages}
  {25P} (\bibinfo {year} {1992})}\BibitemShut {NoStop}%
\bibitem [{\citenamefont {Blokland}\ \emph {et~al.}(2005)\citenamefont
  {Blokland}, \citenamefont {van~der Swaluw}, \citenamefont {Keppens},\ and\
  \citenamefont {Goedbloed}}]{Blokland:2005id}%
  \BibitemOpen
  \bibfield  {author} {\bibinfo {author} {\bibfnamefont {J.~W.~S.}\
  \bibnamefont {Blokland}}, \bibinfo {author} {\bibfnamefont {E.}~\bibnamefont
  {van~der Swaluw}}, \bibinfo {author} {\bibfnamefont {R.}~\bibnamefont
  {Keppens}}, \ and\ \bibinfo {author} {\bibfnamefont {J.~P.}\ \bibnamefont
  {Goedbloed}},\ }\href@noop {} {\bibfield  {journal} {\bibinfo  {journal}
  {Astron. Astrophys.}\ }\textbf {\bibinfo {volume} {444}},\ \bibinfo {pages}
  {337} (\bibinfo {year} {2005})}\BibitemShut {NoStop}%
\bibitem [{Note1()}]{Note1}%
  \BibitemOpen
  \bibinfo {note} {Note that the local shearing wave equations are actually
  nonlinearly valid due to fortuitous cancellations in Eqs.~\protect \textup
  {\hbox {\mathsurround \z@ \protect \normalfont (\ignorespaces \ref
  {shearMHD}\unskip \@@italiccorr )}} upon insertion of the shearing wave
  ansatz.}\BibitemShut {Stop}%
\bibitem [{\citenamefont {Pessah}\ and\ \citenamefont
  {Psaltis}(2005)}]{Pessah:2005cp}%
  \BibitemOpen
  \bibfield  {author} {\bibinfo {author} {\bibfnamefont {M.~E.}\ \bibnamefont
  {Pessah}}\ and\ \bibinfo {author} {\bibfnamefont {D.}~\bibnamefont
  {Psaltis}},\ }\href@noop {} {\bibfield  {journal} {\bibinfo  {journal}
  {Astrophys. J.}\ }\textbf {\bibinfo {volume} {628}},\ \bibinfo {pages} {879}
  (\bibinfo {year} {2005})}\BibitemShut {NoStop}%
\bibitem [{\citenamefont {Kirillov}\ and\ \citenamefont
  {Stefani}(2013)}]{Kirillov:2013db}%
  \BibitemOpen
  \bibfield  {author} {\bibinfo {author} {\bibfnamefont {O.~N.}\ \bibnamefont
  {Kirillov}}\ and\ \bibinfo {author} {\bibfnamefont {F.}~\bibnamefont
  {Stefani}},\ }\href@noop {} {\bibfield  {journal} {\bibinfo  {journal} {Phys.
  Rev. Lett.}\ }\textbf {\bibinfo {volume} {111}},\ \bibinfo {pages} {061103}
  (\bibinfo {year} {2013})}\BibitemShut {NoStop}%
\bibitem [{\citenamefont {Bondeson}\ \emph {et~al.}(1987)\citenamefont
  {Bondeson}, \citenamefont {Iacono},\ and\ \citenamefont
  {Bhattacharjee}}]{Bondeson:1987bq}%
  \BibitemOpen
  \bibfield  {author} {\bibinfo {author} {\bibfnamefont {A.}~\bibnamefont
  {Bondeson}}, \bibinfo {author} {\bibfnamefont {R.}~\bibnamefont {Iacono}}, \
  and\ \bibinfo {author} {\bibfnamefont {A.}~\bibnamefont {Bhattacharjee}},\
  }\href@noop {} {\bibfield  {journal} {\bibinfo  {journal} {Physics of
  Fluids}\ }\textbf {\bibinfo {volume} {30}},\ \bibinfo {pages} {2167}
  (\bibinfo {year} {1987})}\BibitemShut {NoStop}%
\bibitem [{Note2()}]{Note2}%
  \BibitemOpen
  \bibinfo {note} {These ordering assumptions arise from assuming the solution
  varies faster than the background equilibrium. If either $\protect
  \mathaccentV {bar}016{\nu }$ or $\protect \mathaccentV {bar}016{\eta }$ are
  very large one may wish to alter these, leading to slightly different
  shearing wave equations}\BibitemShut {NoStop}%
\bibitem [{Note3()}]{Note3}%
  \BibitemOpen
  \bibinfo {note} {The lack of an azimuthal field renders Eqs.~\protect \textup
  {\hbox {\mathsurround \z@ \protect \normalfont (\ignorespaces \ref
  {SWeqns}\unskip \@@italiccorr )}} and \protect \textup {\hbox {\mathsurround
  \z@ \protect \normalfont (\ignorespaces \ref {GlobalSWeqns}\unskip
  \@@italiccorr )}} identical in this example}\BibitemShut {NoStop}%
\bibitem [{Note4()}]{Note4}%
  \BibitemOpen
  \bibinfo {note} {In fact, a similar result holds for short time growth in the
  equivalent sheared rotating hydrodynamic system, where it seems that there is
  never a transition to the turbulent state}\BibitemShut {NoStop}%
\bibitem [{\citenamefont {Farrell}\ and\ \citenamefont
  {Ioannou}(1994)}]{Farrell:1994wh}%
  \BibitemOpen
  \bibfield  {author} {\bibinfo {author} {\bibfnamefont {B.~F.}\ \bibnamefont
  {Farrell}}\ and\ \bibinfo {author} {\bibfnamefont {P.~J.}\ \bibnamefont
  {Ioannou}},\ }\href@noop {} {\bibfield  {journal} {\bibinfo  {journal} {Phys.
  Rev. Lett.}\ }\textbf {\bibinfo {volume} {72}},\ \bibinfo {pages} {1188}
  (\bibinfo {year} {1994})}\BibitemShut {NoStop}%
\end{thebibliography}

%

\end{document}